# DYNAMICS OF ELECTRONS IN GRADIENT NANOSTRUCTURES
# ( EXACTLY SOLVABLE MODEL) .

*A. Shvartsburg,[1] V. Kuzmiak [2] and G. Petite [3]*

[1] Joint Institute of High Temperatures, Russian Acad. of Sciences,
 Izorskaya 13/19, 125412, Moscow, Russia.
[2] Institute of Photonics and Electronics, Czech Acad. of Sciences, v.v.i.
 Chaberska 57, 182 51 Praha 8, Czech Republic.
[3] Laboratoire des Solides Irradies, Ecole Polytechnique, CEA – DSM, CNRS,
 91128, Palaiseau, France.

**Abstract.**

A flexible multi-parameter exactly solvable model of potential profile, containing an arbitrary number of continuous smoothly shaped barriers and wells, both equal or unequal, characterized by finite values and continuous profiles of the potential and of its gradient, is presented. We demonstrate an influence of both gradient and curvature of these potentials on the electron transport and spectra of symmetric and asymmetric double–well (DW) potentials. The use of this model is simplified due to one to one correspondence between the algorithms of calculation of the transmittance of convex barriers and energy spectra of concave wells. We have shown that the resonant contrast between maximum and minimum in over-barrier reflectivity of curvilinear barrier exceeds significantly the analogous effect for rectangular barrier with the same height and width. Reflectionless tunneling of electrons below the bottom of gradient nanostructures forming concave potential barriers is considered. The analogy between dynamics of electrons in gradient fields and gradient optics of heterogeneous photonic barriers is illustrated.

# I - Introduction.

The ability to tailor the potential of electrons on the scale of their de Broglie wavelength has opened the new horizons in nanoelectronics. Dynamics of quantum particles in these heterogeneous fields, shaped by continuous spatial variations of potential as well as its gradient, attracts a growing attention in several fields of atomic, optical and solid state physics. Namely, engineering of complicated potential barriers for controlled transport of electrons in semiconductor superlattices and heterostructures /1-3/, is widely used in microelectronic systems. This approach, generalized for traveling and tunneling regimes in motion of quasiparticles, proves to be the useful tool for analysis of the dynamics of polaritons in molecular crystals /4/ as well as quantum defects /5/ and magnetic moments /6/ in solids. A special attention was brought to periodical potentials, particularly to the dynamics of atom wavepackets in magnetic potentials, supported by current – carrying wires /7/ and, in particular, to the control of atomic ensembles and matter waves in optical lattices, arising from a set of interfering laser beams /8-10/. A wealth of literature has been devoted to transport and trapping of quantum objects in the double-well (DW) potentials of both natural and technological origin /11-13/.

By analogy with gradient optics, dealing with the propagation of electromagnetic waves through heterogeneous photonic barriers /14/, we will consider here the dynamics of de Broglie waves in gradient nanostructures, characterized by variety of smoothly shaped potential profiles. Since the characteristic spatial scales of potentials discussed are comparable with the de Broglie wavelength of a quantum object, the perturbative approach or WKB approximation fail in such cases, and exact analytical solutions of the Schrödinger equation are in need. A few well known exactly solvable potentials were pioneered as long ago as in the first years of quantum mechanics /15/. The exact analytical results for scattering on periodical and DW potentials were restricted to models represented by sequences of rectangular boxes /16/, chains of coordinate $\delta(z-z_n)$ functions /17/ and combinations of rectangular and linear barriers (the "trapezoidal" profile) /18/. Smoothly shaped wells and barriers of potential profiles in realistic quantum structures were approximated in these simple models by broken lines. Herein these approximations result in the appearance of unphysical corners and infinite derivatives of the profile, bringing distortions in the obtained electron spectra. Moreover, such models have no additional free parameters, permitting to link the electron spectra with the shape and symmetry of curvilinear tops and bottoms of realistic potential profiles. Another model of DW profile, presented by "crossing parabolas" /19/, results in an unphysical sharpening at the crossing point of parabolas, corresponding to the important area of tunneling. Attempts to improve these results by means of empirical "rounding" of corners had revealed an essential dependence of spectra upon the method of rounding.

In contrast, we present in this paper an analysis of scattering and trapping of electrons in the framework of exactly solvable 1D multi-parameter model of smoothly shaped potential U(z). To avoid any uncertainty, connected with tangent of piece-wise parts of this potential, our model is characterized by continuity of both the profile U(z) and its gradient grad U(z). The building blocks of this model are convex and concave arcs, n-th arc $U_n(z)$ being defined on the segment $0 \leq z_n \leq d_n$ as

$$U_n(z) = U_0\left(1 - \frac{1}{g_n} + \frac{W_n^2(z)}{g_n}\right); W_n(z) = \left[\cos(z/L_n) + M_n \sin(z/L_n)\right]^{-1} \quad (1)$$

$$W(z=0) = W(z=d) = 1$$

Here $U_0$ is some normalization constant with the dimension of energy. The number of such arcs may be arbitrary; each concave and convex arc has smooth contact points with the neighbouring segments at the level $U(z) = U_0$. The potential inside the n-th segment, the value $U_0$ being chosen, is presented by formula (1), containing three free parameters $g_n$, $M_n$ and $L_n$; where $g_n$ and $M_n$ are dimensionless constants, and the characteristic scales $L_n$ have the dimension of length. Positive (negative) values of $g_n$ relate to the concave (convex) arcs with minima (maxima) of potential

$$U_{n(\min,\max)} = U_0\left[1 \mp \frac{M_n^2}{g_n(1 + M_n^2)}\right] . \quad (2)$$

In a limiting case $M = 0$, $g = 1$ model (1) reduces to $U(z) = U_0[\cos(z/L)]^{-2}$; this limit, unlike (1), contains only one free parameter $L$, and was used in /20/ for the analysis of polarization-dependent tunneling of light through gradient dielectric layers.

Normalized potentials (1) are presented on Fig. 1a and 1b. The central peak on Fig. 1a, located between the points $z = 0$ and $z = d$, is surrounded by two concave half-arcs, corresponding to the segments $-0.5d \leq z \leq 0$ and $d \leq z \leq 1.5d$. The potential in the range $z \leq -0.5d$ and $z \geq 1.5d$ is assumed to be equal to $U_{\min}$ (2). Analogously, the well on Fig.1b, located in the segment $0 \leq z \leq d$, is surrounded by two convex half-arcs, corresponding to the segments $-0.5d \leq z \leq 0$ and $d \leq z \leq 1.5d$ with the potential in the range $z \leq -0.5d$ and $z \geq 1.5d$ equal to $U_{\max}$. Thus, the potential curves are continuous at all the characteristic points $z = -0.5d$, $z = 0$, $z = d$ and $z = 1.5d$. Free parameters $g$, $M$ and $L$ for each arc are determined by the values of $U_{min,max}$ and distance $d$ between the neighbouring points $U = U_0$:

$$\frac{d}{L} = Arc\cos\left(\frac{1 - M^2}{1 + M^2}\right). \tag{3}$$

To provide positive values of $(U_n)_{min}$ we will consider the concave arcs with $|g_n| > M_n^2/(1 + M_n^2)$. The demand of equal tangents of adjacent n-th and (n + 1)-th arcs at the contact points $U_0 = 1$ results in condition

$$\frac{M_n}{g_n L_n} = -\frac{M_{n+1}}{g_{n+1} L_{n+1}}. \tag{4}$$

Combination of several concave and convex arcs, obeying to condition (4), can present smooth double-wells as well as periodical profiles, shown, e.g., on Fig.2 - 6.

The paper is organized as follows: In Section II we describe the exact analytical solution of Schrödinger equation for smoothly shaped wells and barriers of potential (1) and the relevant boundary conditions. The multiparameter flexibility of boundary conditions is illustrated in Section III on the simplest example - calculation of electron transport through gradient barriers of finite width, formed from the arcs (1). An important effect of reflectionless tunneling (complete transmission) of electrons through a gradient potential barrier is considered in Section IV. We describe in Section V a simple standardized algorithm, based on the same approach, for calculation of spectra of DW potentials, both symmetric and asymmetric. Some generalizations of these results are summarized in Section VI. The symmetry properties of the obtained formulae, simplifying the calculations, are presented in the Appendices 1 and 2.

**II - Eigenfunctions of smoothly shaped multiparameter potentials.**

To solve the Schrödinger equation

$$\frac{d^2\psi_n}{dz^2} + \frac{2m}{\hbar^2}(E - U_n)\psi_n = 0 \tag{5}$$

for the potential U (1) let us introduce the new function f and new variable $\eta$

$$f_n = \psi_n \sqrt{W_n(z)} \; ; \quad \eta_n = \int_0^z W_n(z')dz' \tag{6}$$

$$\eta = \frac{L}{\sqrt{1+M^2}} \ln\left[\frac{1 + m_+ tg(z/2L)}{1 - m_- tg(z/2L)}\right]; \; m_\pm = \sqrt{1+M^2} \pm M \; . \tag{7}$$

The variable $\eta$ as well as parameters M, $m_+$ and $m_-$ have to be taken in fact for each n-th interval as $\eta_n$, $M_n$, $(m_+)_n$ and $(m_-)_n$; however, for the sake of the simplicity of notation we omit the index n hereafter. We note the useful property of quantities $m_\pm$: $m_+ m_- = 1$. The factor W(z) in (1) can be expressed in terms the new variable $\eta$ :

$$W(z) = \frac{ch\left[\frac{\eta}{L}\sqrt{1+M^2} - ArcshM\right]}{\sqrt{1+M^2}} \; ; \; ArcshM = \ln(m_+) \; . \tag{8}$$

By introducing the normalized variable x

$$x = \frac{\eta}{L}\sqrt{1+M^2} - \ln(m_+) = \ln\left[\frac{m_- + tg(z/2L)}{1 - m_- tg(z/2L)}\right] \quad (9)$$

and substituting (6) and (8) into the Schrödinger Eq. (5), we obtain the master equation, governing the function f:

$$\frac{d^2 f}{dx^2} - f\left(q^2 + \frac{\Lambda}{ch^2 x}\right) = 0. \quad (10)$$

The coefficients of Eq. (10) are expressed in terms of the parameters of potential

$$q_{\pm}^2 = \frac{1}{4} \pm \frac{L^2}{a^2} \frac{1}{g(1+M^2)} \;;\; \Lambda = \frac{1}{4} - \frac{L^2}{a^2}\theta \;;\; \theta = \varepsilon - 1 + \frac{1}{g} \;;\; \varepsilon = \frac{E}{U_0}, \quad (11)$$

where $a$ is the quantum spatial scale

$$a = \frac{\hbar}{\sqrt{2mU_0}}. \quad (12)$$

Let us point out, that the value of parameter $g$ in (11) for well (barrier) range is positive (negative). Thus, the equations for both well and barrier range are presented in similar forms. This similarity simplifies the forthcoming analysis. To find the solutions of Eq. (10) we introduce a new function $F$ and a new variable $u$:

$$u = \frac{1}{2}(1 - thx) \; ; \; F = f(chx)^{-q} . \tag{13}$$

Owing to transform (13) Eq. (10) is reduced to a standard form of hypergeometric differential equation /21/

$$u(1-u)\frac{d^2F}{du^2} + \left[\gamma - (1+\alpha+\beta)u\right]\frac{dF}{du} - \alpha\beta F = 0 \tag{14}$$

$$\alpha, \beta = \frac{1}{2} - q \pm \frac{L}{a}\sqrt{\theta} \; ; \; \gamma = 1 - q . \tag{15}$$

The hypergeometric equation (14) is known to have two linearly-independent solutions. Since the parameters $\alpha, \beta, \gamma$ are linked by the correlation $Re((\alpha + \beta + 1) = 2\gamma$, these solutions are given by hypergeometric functions $F_1$ and $F_2$:

$$F_1 = F(\alpha, \beta, \gamma, u) \; ; \; F_2 = F(\alpha, \beta, \gamma, 1-u). \tag{16}$$

Moreover, since $Re(\alpha + \beta - \gamma) = -q < 0$, the hypergeometric series $F_1$ and $F_2$ are absolutely converging within the circle $|u| = 1$ /21/. Finally, by combining the expressions (6), (13) and (16), we will obtain the general solution of Schrödinger Eq. (4) for each well and bottom of periodical potential in the form

$$\psi = A(chx)^{q-\frac{1}{2}}(F_1 + QF_2). \tag{17}$$

Here A is the normalization constant, the values Q have to be defined from the conditions of continuity of logarithmic derivatives $\frac{1}{\Psi}\frac{d\psi}{dz}$ at the points of tangent of the different parts of the potential.

To use the continuity conditions for the wave function one has to determine the values of variables $x$ (9), $u$ (13) and $1-u$ at the points of contact $U(z) = U_0$; thus for profiles shown on Fig.1a and Fig.1b:

$$z = 0 : x = -\ln(m_+) \; ; u = v_+ = \frac{1}{2}\left(1 + \frac{M}{\sqrt{1+M^2}}\right); v_- = 1 - u$$

$$z = d : x = \ln(m_+) \; ; u = v_- \; ; 1 - u = v \qquad (18)$$

$$z = -d/2 \; ; z = 3d/2 : u = 1 - u = v_0 = 1/2$$

These values will be used below for both scattering and eigenvalues problems.

## III - Transport of electrons through convex gradient barrier.

To present the algorithm for calculation of scattering of electrons on the gradient structure one can at first examine the simplest case – reflection of electrons on a single barrier, shown on Fig. 1a. Let us consider an electron with energy E > $U_{min}$ (2), incidenting from z = - ∞ at the point z = - 0.5d. It is convenient to use the continuity conditions in the consecutive order, starting from the right edge of the structure and moving to the left. The electron wave function in the range of constant potential z ≥ 1.5d reads as

$$\psi = B\exp\left[ik(z-1.5d)\right]; k = \frac{1}{a}\sqrt{\varepsilon - 1 + \frac{M^2}{|g|(1+M^2)}}. \qquad (19)$$

Here parameter a is defined in (12), $\varepsilon$ is the normalized energy (11). The logarithmic derivative of function $\Psi$ (19) in the point z=1.5d is equal to ik. By denoting the functions F corresponding to the well as $F_-$, one can find the logarithmic derivative of function $\Psi$ (19) at this point, related to the value of variable v = $v_0$ = ½:

$$\frac{1}{\psi}\frac{d\psi}{dz}\bigg|_{z=1.5d} = -\frac{1}{F_-}\frac{F'_-(v_0)(1-Q_2)}{F_-(v_0)(1+Q_2)}. \qquad (20)$$

By equating the left and right logarithmic derivatives of wave function at the point z = 1.5d, one can find the parameter $Q_2$:

$$Q_2 = -\frac{2ikl + Y_-}{2ikL - Y_-} \; ; \; Y_- = \frac{F'_-(v_0)}{F'(v_0)} . \tag{21}$$

Moving to the left, we will one can evaluate the logarithmic derivatives of wave functions at the point z = d:

$$\left. \frac{1}{\psi} \frac{d\psi}{dz} \right|_{z=d_+} = \frac{D_3}{2L} \; ; \; D_3 = \frac{P_5 + Q_2 P_6}{F_-(v_+) + Q_2 F_-(v_-)} \tag{22}$$

$$\left. \frac{1}{\psi} \frac{d\psi}{dz} \right|_{z=d_-} = \frac{D_2}{2L} \; ; \; D_2 = \frac{P_3 + Q_1 P_4}{F_-(v_+) + Q_1 F_-(v_-)} . \tag{23}$$

All the quantities $P_n$ are collected in Appendix 1 in order to demonstrate the symmetry of their structure. By using the value $Q_2$ from (21), one can calculate the quantity $D_3$ (22); then, using the equality of derivatives (22) and (23), one can determine the value $Q_1$:

$$D_3 = -D_2 \; ; \; Q_1 = -\frac{P_3 + D_3 F_+(v_-)}{P_4 + D_3 F_+(v_+)} . \tag{24}$$

We repeat the same procedure with derivatives of wave function near by the point z = 0. By expressing the right and left derivatives at this point in terms of the $D_1$ and $D_0$ respectively, we will determine $D_1$ and $D_0$ by analogy with formulae (22) – (23):

$$D_1 = \frac{P_1 + Q_1 P_2}{F_+(v_+) + Q_1 F_+(v_-)} \; ; \; D_0 = \frac{P_{-1} + Q_0 P_{-2}}{F_-(v_-) + Q_0 F_-(v_+)} . \tag{25}$$

Calculation of $D_1$ by means of $Q_1$ yields the value $Q_0$:

$$D_1 = -D_0 \;;\; Q_0 = -\frac{P_{-1} + D_1 F_-(v_-)}{P_{-2} + D_1 F_-(v_+)}. \tag{26}$$

While calculating the value $Q_0$ we used the model $W(z)$ (1) and solution (17) in the range $z \leq 0$, replacing in the relevant formulae $M \to -|M|$; where the coordinate $\eta$ (6) becomes negative, and the normalized variable x (9) reads as $x = \frac{\eta}{L}\sqrt{1+M^2} + ln(m_+)$.

Now one can obtain the complex reflection coefficient R by using the continuity condition at the left boundary $z = -d/2$. The wave function at the range of constant potential $z \leq -d/2$ reads as $\Psi_1 = B_1 \exp[ik(z + 0.5d)]$. By calculating the derivative of wave function $\Psi$ (17) at the point $z = -d/2 + 0$ ( $v = v_0 = ½$) by analogy with (20), one obtains

$$R = \frac{2ikL(1+Q_0) + Y_-(1-Q_0)}{2ikL(1+Q_0) - Y_-(1-Q_0)}, \tag{27}$$

where the dimensionless parameter $Y_-$ was defined in (21).

Thus, the calculation of reflection coefficient R for some energy E can be performed according to the following standardized procedure:

1. To evaluate $Q_2$ from (21).

2. To determine $D_3$ by substituting $Q_2$ into (22).

3. To calculate $Q_1$ (24) by using $D_3$.

4. To determine $D_1$ by substituting $Q_1$ into (25).

5. To calculate $Q_0$ (26) by using $D_1$.

6. Finally, to evaluate R by substituting $Q_0$ into (27).

This procedure can be presented symbolically by chain

$$Q_2 \to D_3 \to Q_1 \to D_1 \to Q_0 \to R. \qquad (28)$$

Reflectance of more complicated structures can be examined in a similar fashion. Namely, to find the reflectance of two similar barriers one can start again from the right edge, located now at the point $z = 3.5d$, we have to start from $Q_4$, given by Eq. (21) due to replacement of $Q_2$ by $Q_4$. Then, using the continuity conditions at $z = 3d$, one obtains $D_7$ while the values $D_7 - D_4$ are given in Appendix 1 and the value $Q_3$ is given by condition $D_7 = -D_6$:

$$Q_3 = -\frac{P_{11} + D_5 F_+(v_-)}{P_{12} + D_7 F_+(v_+)}. \qquad (29)$$

The substitution of $Q_3$ to $D_5$ and using the continuity condition $D_5 = -D_4$ at the point $z = 2d$ yields the value $Q_2$:

$$Q_2 = -\frac{P_7 + D_5 F_-(v_-)}{P_8 + D_5 F_-(v_+)}. \qquad (30)$$

Using this value of $Q_2$ in the chain (28), one can calculate the reflection coefficient R for the structure, containing two peaks; the algorithm of calculation of R in this case can be represented by the following sequence of operations by generalizing the chain (28):

$$Q_4 \to D_7 \to Q_3 \to D_5 \to Q_2 \to D_3 \to Q_1 \to D_1 \to Q_0 \to R. \tag{31}$$

Transmittance of potential barriers $|T|^2$ can be found as

$$|T|^2 = 1 - |R|^2. \tag{32}$$

The forthcoming generalization of this approach for a structure containing an arbitrary amount of alternating peaks and wells can be performed analogously.

Transmittance for electrons with energy E propagating through a single gradient barrier(Fig. 1a) is shown in Fig. 7 (curve 1). The effect of the barrier form-factor (Fig. 1a) is demonstrated by means of the transmittance of rectangular barrier with the same width $d_0 = 2d$ and with the same potential minima and maxima $U_{min}$ and $U_{max}$ – see curve 2 in Fig. 7. In the case $E \leq U_{max}$, $w = E/U_{max} \leq 1$ the transmittance of the rectangular barrier can be written as

$$|T|^2 = \frac{4w(1-w)(w-\sigma)}{4w(1-w)(w-\sigma) + (1-\sigma^2)\text{sh}^2(qd_0)}$$

$$\sigma = \frac{U_{min}}{U_{max}}; \quad q = \frac{\sqrt{1-w}}{b}; \quad b = \frac{\hbar}{\sqrt{2mU_{max}}} \tag{33}$$

To use (33) in a case ($E > U_0$, $w > 1$) one has to replace q in Eq. (33) by $i\sqrt{w-1}/b$ and sh($qd_0$) by $i \sin(\sqrt{w-1}d_0)/b$. Let us note that, due to difference between $U_0$ and $U_{max}$, the characteristic scale b (33) is distinct from scale a (11), $b = ap$

$$p = \sqrt{\frac{U_0}{U_{max}}}. \tag{34}$$

To compare the transmittance of both barriers for the same energies, one has to keep in mind that the normalized energy for gradient barrier $\varepsilon$ (11) is related to the normalized energy for a rectangular barrier $w$ as $\varepsilon = wp^{-2}$.

Let us stress out the following peculiarities of the graphs $|T(w)|^2$, determining the electron transport processes:

1. Transmittance of rectangular barrier is connected with discontinuities of potential $U(z)$ and its gradient at the boundary points $z = -0.5\,d$ and $z = 1.5\,d$, meanwhile in the case of the gradient barrier profile $U(z)$ is continuous as well as its gradient; however, the transmittance of gradient barrier is influenced by the discontinuities of curvature of $U(z)$ at these boundary points.

2. The transmittances depicted in Fig. 7 reveal the minima for energies exceeding the maxima of barriers($w = 1$). These minima correspond to the maxima of the over-barrier reflection coefficients $|R|^2 = 1 - |T|^2$. This over-barrier reflection which is associated with the resonant correlations between the de-Broglie wavelength and the effective thickness of the barrier corresponding to gradient barrier is much stronger than that of corresponding to the box-like barrier. Specifically, the first maximum of the reflection coefficient for gradient barrier(curve1) is $|R_1|^2{}_1 = 0.282$, while for box-like barrier $|R_1|^2{}_2 = 0.136$. The second maximum of the reflectance for gradient barrier slightly differs from the first one: $|R_2|^2{}_1 = 0.24$, while the same coefficient for box-like barrier is extremely small: $|R_2|^2{}_2 = 0.019$. The energy of electron corresponding to resonant over-barrier reflection from box-like barrier found from (33) is inversely proportional to barrier width $d_0$. By using this correlation qualitatively for gradient barrier, whose effective width in the case at hand is several times smaller than $d_0$, one can expect large values of energies corresponding to the maxima of over-barrier reflection from gradient barrier – see Fig. 7: $w_1 = 3.5$, $w_2 = 5.95$. Thus, the gradient barrier can possess the filtering properties for transport of electrons with some over-barrier energies.

3. The value of $|T|^2$ for gradient profile (Fig. 7, curve 1) can be used for analysis of transmittance of other profiles, obtained from those discussed ones by means of special transform. Inspection of formulae (11) shows that the master equation (10), governing the wave function, and its solutions remain unchanged, when parameters $M$ and $g$ are fixed, while the electron energy $E$, potential $U_0$ and scale $L$ can vary in such a way that the ratios $\varepsilon = E/U_0$ and $L/a$ also remain unchanged. Here the parameters $Q_h$ and $D_h$ (28) are invariant and thus the values of reflection coefficient $R$ as well as $|T|^2$ are invariant too. Therefore, by characterizing the coupled variations of quantities $E$, $U_0$ and $L$ by parameter $h$ one can see that the single gradient barriers $U_1$ and $U_2$ with normalization potentials $(U_2)_0$ and $(U_1)_0$, linked by relations

$$M_1 = M_2 \,;\; g_1 = g_2 \,;\; L_2 = L_1 h^{-1} \,;\; (U_2)_0 = (U_1)_0 h^2 \qquad (35)$$

provide equal transmittances for electrons with energies $\varepsilon_1$ and $\varepsilon_2 = \varepsilon_1 h^{-2}$, while the widths of these profiles are also correlated: $d_2 = d_1 h^{-1}$. Such potential profiles, corresponding to different values of $h$ and providing equal transmittance for electrons with such energies $\varepsilon_1$ and $\varepsilon_2$ are depicted in Fig. 8a.

It is worth to compare the relations (35), obtained for potential (1), with the relation between the energy levels E of box-like potential and its width d: the product $Ed^2$ is known to remain constant for a given quantum number. Here the width d can be changed independently of the potential maximum, conserving the box-like potential shape. Unlike the latter, relations (35) present the coupled transform of scale L and

potential parameter $U_0$, resulting in profound reshaping of potential profile shown in Fig. 8a

## IV - Reflectionless tunneling of electrons through a concave potential barrier.

Tunneling of electrons with energy $E$ through a box-like potential barrier with height $U_{max} > E$, described by Eq. (33), is characterized by a transmittance which is always smaller than unity; therefore the reflection coefficient is non-zero. However, this situation can be profoundly different for electrons tunneling through a concave potential barrier with minimum $U_{min}$. Namely, for some energies smaller than $U_{min}$ a peculiar regime of reflectionless tunneling ($|R|^2 = 0$) proves to be possible. This regime arise from the interference of forward and backward electronic de Broglie waves inside the barrier.

To visualize the underlying physics of this effect let us consider the simplest geometry of gradient barrier, formed by several adjacent concave arcs with equal parameters M, g, L and $U_0 = U_{max}$ at the top of the base labeled by $U_p$ (Fig. 9). Rigorously speaking, it is necessary to smooth out the discontinuities of gradient U at the points $U = U_{max}$ by assuming existence a small intermediate layer formed by convex arc with parameters $M_1 << 1$, $L_1 << L$ and $g_1 = g$, where the condition (4) reads as $M_1/M = L_1/L$. Since both the width of this intermediate layer (3) $d_1 = 2L_1M_1 << d$ and its relative height $(U_1)_{max}/U_0 - 1 = M_1^2/g << 1$ are small, and, moreover, the tunneling particle energy $\varepsilon$ is smaller than the barrier minimum ($U_0 > U_{min} > \varepsilon$), one can neglect the influence of this layer on tunneling, considering the reflection of particle on the discontinuities of grad U at the boundaries $z = 0, d, 2d$.

To avoid a tedious algebra, connected with Eq. (10), we consider a special case $\Lambda = 0$. In this case solution of Eq. (10) is expressed in terms of the elementary functions $\exp(\pm qx)$. The wave function $\Psi$ (17) inside the barrier can be written by means of variable x (9) and parameter q (10):

$$\Psi = A(chx)^{q-\frac{1}{2}}[\exp(-qx) + Q\exp(qx)] . \tag{36}$$

Representing the electron incidenting from the left on the boundary z = 0 by means of traveling wave

$$\Psi_{inc} = B\exp(ipz); \quad p = \frac{\sqrt{\varepsilon - \delta}}{a}; \quad \delta = \frac{U_p}{U_0} \tag{37}$$

one can at first examine the reflection coefficient R for tunneling through one barrier between the points z=0 and z = d. The values of variable x and the functions in (36) at these points are

$$x|_{z=0} = x_0 = -\ln(m_+); \quad th(x_0) = -\frac{M}{2\sqrt{1+M^2}} = -\ell \; ; \; x|_{z=d} = -x_0 . \tag{38}$$

The continuity condition for logarithmic derivative $\frac{1}{\Psi}\frac{d\Psi}{dz}$ at z = 0 yields the equation, governing the reflection coefficient R

$$\frac{ip(1-R)}{1+R} = \frac{\sqrt{1+M^2}}{L}\left\{\ell - \frac{q[\exp(-qx_0) - Q\exp(qx_0)]}{\exp(-qx_0) + Q\exp(qx_0)}\right\} ; \; \ell = \frac{M}{2\sqrt{1+M^2}} . \tag{39}$$

The unknown quantity Q in (39) is defined from the continuity condition at z = d:

$$Q = \frac{(1+\chi)\exp(2qx_0)}{1-\chi} \; ; \; \chi = \frac{1}{q}\left(\ell + \frac{ipL}{\sqrt{1+M^2}}\right). \tag{40}$$

Substitution of Q (40) into Eq. (39) yields the expression for the complex reflection coefficient R:

$$R = \frac{\left(q^2 + \ell^2 + \frac{(pL)^2}{1+M^2}\right)th(2qx_0) + 2q\ell}{\left(q^2 + \ell^2 - \frac{(pL)^2}{1+M^2}\right)th(2qx_0) - 2q\ell - \frac{2ipL}{\sqrt{1+M^2}}[q + \ell\, th(2qx_0)]} \tag{41}$$

Formula (41) is the main result of this Chapter. By using the value $x_0$ (38) one can rewrite the term $th(2qx_0)$ in the form

$$th(2qx_0) = \frac{1 - (m_+)^{2q}}{1 + (m_+)^{2q}}. \tag{42}$$

It is remarkable, that coefficient R for the concave barrier (Fig. 9), unlike the reflection coefficient for box-like barrier, can reach the zero value R = 0. The condition of this nullification (reflectionless tunneling) follows from (41):

$$\frac{(m_+)^{2q} - 1}{(m_+)^{2q} + 1} = \frac{4qM\sqrt{1+M^2}}{M^2 + 4q^2(1+M^2) + 4(pL)^2}. \tag{43}$$

By proceeding in a similar manner, we one obtains the condition of reflectionless tunneling through the system of n contiguous concave barriers following from (43) due to replacement

$(m_+)^{2q} \to (m_+)^{2qn}$.

An example of such barrier, shown on Fig. 9, can be constructed, e.g., by means of concave arc with parameters M = 2.02 and g = 1.35, L = 0.325 nm, $U_0$ = 1eV. Thus, for the potential minimum $U_{min}/U_0$ = 0.4 and base height $\delta$ =0.3 the reflectionless tunneling appears for electrons with energy $\varepsilon$ = 0.35. Here the condition $\Lambda$ = 0, simplifying the master equation (10) is satisfied, and the energy $\varepsilon$ is located between the minimum and the base.

This unusual quantum phenomenon of total transparency ( $|R|^2$ = 0, $|T|^2$ = 1) of gradient potential barrier for electrons with energy E, tunneling through the forbidden zone of this barrier E < $U_{min}$, illustrates a key role of gradient and curvature of potential profile U(z) on reflectance/transmittance spectra of barrier. This phenomenon does not occur for the transparency $|T|^2$ of box-like potential (33), when the equation $|T|^2$ = 1 has no solutions. Treating the total transparency as a reflectionless tunneling of de Broglie waves, one can emphasize the analogy of this quantum effect with the classical wave effect - reflectionless tunneling of electromagnetic waves through gradient photonic barriers /22/. Both effects represent new phenomena associated with the effective transmission of particles and waves through non-transparent media.

## V - Spectra of double – well potentials.

Spectra of electron energy $\varepsilon_n$ in the continuously shaped double-well (DW) potential are important for study of condensed matter systems /23/ and quantum information processing /24/. A particular interest is stimulated by the perspectives of controlled manipulation of ultracold neutral atoms by means of their spin-dependent motion in DW potential, formed by optical lattices /25, 26/. Such spectra can be found by means of the formalism developed above. For simplicity let us examine firstly an auxiliary problem – the spectrum of single-well potential, formed by one concave arc, surrounded by two convex half–arcs (1); with the profile placed in the segment $-d/2 \leq z \leq 1.5d$ (Fig. 1b). Starting again from the right side of this structure one can present the wave function of confined electron in the range $z \geq 1.5d$ in the form

$$\Psi = B \exp[-\chi(z - 1.5d)] \; ; \; \chi = \sqrt{1 + \frac{M^2}{|g|(1+M^2)} - \varepsilon} \, , \tag{44}$$

where

$$1 - \frac{M^2}{|g|(1+M^2)} < \varepsilon < 1 + \frac{M^2}{|g|(1+M^2)} \, . \tag{45}$$

By comparing this problem with the problem of electron scattering at a single peak (1), one can see that the wells and peaks in these problems are interchanged; however, the general solution (19) can be used in this geometry as well. Therefore, replacing the factors Q, describing the interference of forward and backward waves in (19), by factors $\Phi$, one obtains from continuity conditions at the point z =1.5d:

$$\Phi_2 = \frac{Y_+ - 2\chi L}{Y_+ + 2\chi L} \ ; \quad Y_+ = \frac{F'_+(v_0)}{F_+(v_0)} \ . \tag{46}$$

Then, considering the continuity conditions at z = d ($G_3$ = - $G_2$, the quantities G are defined in the Appendix 2) and z = 0 ($G_1$ = - $G_0$) one obtains the parameters $\Phi_1$ and $\Phi_0$:

$$\Phi_1 = -\frac{K_3 + G_3 F_-(v_-)}{K_4 + G_3 F_-(v_+)}; \quad \Phi_0 = -\frac{K_{-1} + G_1 F_+(v_-)}{K_{-2} + G_1 F_+(v_+)} \ . \tag{47}$$

On the other hand, the parameter $\Phi_0$ can be found independently from the continuity conditions at z = - 0.5d. By representing the wave function at z ≤ - 0.5d in the form $\Psi = \exp[\chi(z + 0.5d)]$, one obtains

$$\Phi_0 = \frac{Y_+ + 2\chi L}{Y_+ - 2\chi L} \ . \tag{48}$$

The values of energy $\varepsilon_n$, providing the equality of expressions (47) and (48) for $\Phi_0$, yield the eigenvalues of electron energy $\varepsilon_n$.

Let us point out, that the expression (46) for $\Phi_2$ is transformed to expression (21) for $Q_2$ due to replacements $Y_+ \to Y_-, \chi \to$ -ik. Further, by making the replacements

$$q_{\pm} \to q_{\mp}\ ;\ S_{\pm} \to S_{\mp}\ ;\ P_m \to K_m;\ F_{\pm}(v_{\pm}) \to F_{\mp}(v_{\pm});\ B_{\pm}(v_{\pm}) \to B_{\mp}(v_{\pm});\ \Phi_{3,2,1} \to Q_{3,2,1} \quad (49)$$

one can find the eigenvalues of electron energy $\varepsilon_n$ in the potential under considered, following the scheme of analysis (31) and by using parameters $K_n$ and $G_n$ from Appendix 2. Thus, e.g., taking into account the recursive formula (A.3), we transform the quantities $P_5$, $P_6$ to $K_5$, $K_6$, which are needed for calculation of parameter $G_3$, analogous to $D_3$ (24). The quantities $\Phi_1$ and $\Phi_0$ can be found by using the transforms of $Q_1$ (24) and $Q_0$ (26).

This scheme of computation of $\varepsilon_n$ can be presented symbolically in the form, similar to (28):

$$\Phi_2 \to G_3 \to \Phi_1 \to G_1 \to \Phi_0\ . \quad (50)$$

In the case of the single-well potential shown on Fig. 1b we obtain the following eigenvalues : $\varepsilon_1 = 1.528$ and $\varepsilon_2 = 1.595$. For all such numerical applications, one needs to specify the value of $U_0$. Here, as well as for all the numerical calculations hereunder, $U_0$ was set equal to 1 eV.

By proceeding in a similar fashion one can find the eigenvalues of double-well potential, generalizing the scheme (31):

$$\Phi_4 \to G_7 \to \Phi_3 \to G_5 \to \Phi_2 \to G_3 \to \Phi_1 \to G_1 \to \Phi_0\ . \quad (51)$$

The quantity $\Phi_0$ is given by Eq. (40); on the other hand, this quantity can be calculated by means of sequence (53), where $\Phi_4$ is equal to $\Phi_2$ (46), while the values $\Phi_3$ and

$\Phi_2$, determined for the double-well potential from the continuity conditions at z = 3d ($G_7 = -G_6$) and z = 2d ($G_5 = -G_4$), are:

$$\Phi_3 = -\frac{K_{11} + G_7 F_-(v_-)}{K_{12} + G_7 F_-(v_+)} \; ; \; \Phi_2 = -\frac{K_7 + G_5 F_+(v_-)}{K_8 + G_5 F_+(v_+)} . \tag{52}$$

By comparing the quantities $\Phi_0$, obtained due to calculations (49) for different values of energy $\varepsilon$ from the interval (45), and the quantity $\Phi_0(\varepsilon)$, given by (48), one obtains the values $\varepsilon_n$, providing the equality of both quantities. These values form the discrete spectrum of electron energies $\varepsilon_n$ for double-well potential.

To illustrate the flexibility of this approach, let us find the spectra of several DW potential profiles, distinguished by their geometry. All such profiles, shown on Fig. 2 – 6, are characterized by smooth transition to the surrounding constant potential $U_c$. Namely, the profiles depicted on Fig. 2 have equal depths of both wells, while the central maximum $U_{max}$ may be equal to $U_c$ (Fig. 2,a) or (Fig. 2,b) either larger or smaller than $U_c$. Finally, Fig. 3 presents the asymmetrical potential with $U_{max} = U_c$, with the depth of wells are unequal. Using the approach described above one obtains for these DW profiles the following eigenvalues of normalized electron energy $\varepsilon = E/U_0$:

$\varepsilon_1 = 1.404$, $\varepsilon_2 = 1.418$ (Fig. 2,a) ; $\varepsilon_1 = 1.108$, $\varepsilon_2 = 1.412$ (Fig. 2,b);

$\varepsilon_1 = 0.989$, $\varepsilon_2 = 1.285$ (Fig. 2,b); $\varepsilon_1 = 1.456$; $\varepsilon_2 = 1.522$ (Fig. 3).

Thus including of a second well results in a lowering of the eigenvalues with respect to the surrounding level U = 1.6.

It is remarkable that, using formulae (35), one can transform the continuous well profile $U_1$ into another well profile $U_2$, which differs from $U_1$ by depth $(U_2)_{min} = (U_1)_{min} h^2$, level of surrounding potential $(U_c)_2 = (U_c)_1 h^2$ and width $d_2 = d_1 h^{-1}$. The values M, g and ratio L/a are assumed to remain constant, by analogy with Fig. 8a. Some examples of such transformed wells are shown in Fig. 8b. Here the energy eigenvalues for wells $U_2$ and $U_1$ are linked by the correlation $(\varepsilon_n)_2 = (\varepsilon_n)_1 h^2$. Thus, the energies for transformed potential $(\varepsilon_n)_2$ can be found from $(\varepsilon_n)_1$ without evaluating (46) – (52).

To emphasize the spectral peculiarities of curvilinear wells let us compare their energy levels to that, obtained for the box-like wells with the same values of height $U_{max} - U_{min}$ and width $d_0$ (Fig.1b). By expressing the wave function of trapped electron in the form:

$$z \leq 0 \; ; \; \Psi = \exp(\chi z), \chi = \frac{\sqrt{U_{max} - \varepsilon}}{a} \; ;$$

$$0 \leq z \leq d_0 \; ; \quad \Psi = C[\exp(iqz) + Q\exp(-iqz)]; \; q = \frac{\sqrt{\varepsilon - U_{min}}}{a} \; ; (53)$$

$$z \geq d_0 \; ; \; \Psi = \exp[-\chi(z - d_0)] \; ;$$

and by using the continuity conditions at the boundaries $z = 0$ and $z = d$, one obtains the equation, governing the eigenvalues q:

$$tg(qd) = \frac{2q\chi}{q^2 - \chi^2} \; . \; (54)$$

Here a is the quantum scale (12), C is the normalization constant and distance d is given by Eq.(3). Taking the parameters for box-like potential from Fig.1b, one can find the single root of Eq. (54), that appears in the range $U_{max} > \varepsilon > U_{min}$: $\varepsilon_0 = 0.881$. Unlike the curvilinear potential (Fig. 1b, continuous line), characterized by eigenvalues

$\varepsilon_1$ = 1.528 and $\varepsilon_2$ = 1.595, the box-like potential discussed reveals only one level $\varepsilon_0$, located much lower than both $\varepsilon_1$ and $\varepsilon_2$. Thus, using of box-like model for analysis of continuously-shaped potentials may result in substantial errors in calculation of energy eigenvalues of trapped electrons. Note that a square well with the same total width as the continuously-shaped potential ($d_0=2d$) would exhibit two eigenstates with energies $\varepsilon_1$ = 0.601 and $\varepsilon_2$ = 1.156, so that the increase of the eigenvalues finds is more likely due to the shape of the potential than to its apparent width.

**6. Conclusion**.

In conclusion, we have presented the flexible multiparameter exactly solvable models of continuously shaped 1D quantum mechanical potential barriers and wells U(z), possessing the continuity of both U(z) and grad U(z). By computing the barrier transmittance $|T|^2$ and energy eigenvalues $\varepsilon_n$ for DW potentials, related to different combinations of parameters, we show that both electron transport through these barriers as well as energy spectra of electrons, trapped into SW or DW potentials strongly depend on the gradient and curvature of potential profiles. The standardized successive algorithms for finding $|T|^2$ and $\varepsilon_n$ are presented and one–to–one correspondence of these algorithms to each other is shown. Transforms of profiles U(z), providing the values of $|T|^2$ and $\varepsilon_n$ for transformed potentials from the relevant values for initial potentials without additional calculations, are examined. The effect of reflectionless tunneling through gradient potential barriers, providing the total transmission of electrons with energy, smaller than the minimum of barrier, is shown. The analogy between tunneling of de-Broglie waves through concave potential barrier and tunneling of electromagnetic waves through gradient photonic barrier is emphasized.

For simplicity, the analysis was illustrated above on the example of the simplest geometry of profiles U(z) – single barrier and DW potentials. However, by applying the method in consecutive order the continuity conditions (4), one can consider the flexible models of potential structures, containing an arbitrary amount of barriers and wells. An example of such structure with narrow wells and broad barriers is depicted on Fig. 9. Interchange of parameters $M_1 \leftrightarrow M_2$, $L_1 \leftrightarrow L_2$, $g_1 \leftrightarrow g_2$ yields the structure with narrow peaks and wide barriers. The models of structures (1), containing, unlike these profiles with continuous gradients U(z), the chain of adjacent barriers with discontinuities of gradients of U(z) on the barrier's boundaries, prove to be useful for some fields in the cross-disciplinary physics, e.g., for electromagnetics of transmission lines with continuously distributed parameters /27/.

In the framework the multi-parameter model, one can find the total amount of free parameters of the model discussed, while keeping in mind, that the peak of single barrier $U > U_0$, shown on Fig.1a, is characterized by 4 free parameters: $U_0$, M, g and L. Each of the concave half–arcs, surrounding this peak, the value $U_0$ being fixed, is characterized by three parameters: M, g and L. Taking into account the continuity conditions, imposed at the points z = 0 and z = d, one can assign to this profile 8 free parameters. The flexible models of more complicated structures, containing more concave and convex arcs and, thus, more free parameters, can be considered in the same manner.

Acknowledgements.
One of the authors (A.S.) appreciates the stimulating discussions with Prof. M. Agranat and Prof. V. Fortov. Work of V.K. was partially supported by COST P11 Action.

**Appendix 1.**

These formulae are needed for calculation of electron transport through the multi-parameter barriers.

$P_{-1} = S_+F_-(v_-) + B_-(v_-)$ ; $P_{-2} = S_+F_-(v_+) - B_-(v_+)$ ;

$P_1 = S_-F_+(v_+) - B_+(v_+)$ ; $P_2 = S_-F_+(v_-) + B_+(v_-)$ ;

$P_3 = S_-F_+(v_-) + B_+(v_-)$ ; $P_4 = S_-F_+(v_+) - B_+(v_+)$ ; (A.1)

$P_5 = S_+F_-(v_+) - B_-(v_+)$ ; $P_6 = S_+F_-(v_-) + B_-(v_-)$ ;

$P_7 = S_+F_-(v_-) + B_-(v_-)$ ; $P_8 = S_+F_-(v_+) - B_-(v_+)$ ;

$$S_{\mp} = M(1 - 2q_{\pm}) ; \quad B_{\pm}(v_{\pm}) = \frac{1}{\sqrt{1+M^2}} \left.\frac{dF_{\pm}}{du}\right|_{u=v_{\pm}} ; \quad (A.2)$$

$P_{-1} = P_6$ ; $P_{-2} = P_5$ ; $P_1 = P_4$ ; $P_2 = P_3$ ; (A.3)

$P_m = P_{m+8}$ ; ($m \geq 1$) ;

$$D_4 = \frac{P_7 + Q_2 P_8}{F_-(v_-) + Q_2 F_-(v_+)} ; \quad D_5 = \frac{P_9 + Q_3 P_{10}}{F_+(v_+) + Q_3 F_+(v_-)} ; \quad (A.4)$$

$$D_6 = \frac{P_{11} + Q_3 P_{12}}{F_+(v_-) + Q_3 F_+(v_+)} ; \quad D_7 = \frac{P_{13} + Q_4 P_{14}}{F_-(v_+) + Q_4 F_-(v_-)} .$$

**Appendix 2.**

Parameters $K_m$ and $G_m$ for the eigenvalues problem read as:

$K_{-1} = S_-F_+(v_-) + B_+(v_-)$ ; $K_{-2} = S_-F_+(v_+) - B_+(v_+)$ ;

$K_1 = S_+F_-(v_+) - B_-(v_+)$ ; $K_2 = S_+F_-(v_-) + B_-(v_-)$ ;

$K_3 = S_+F_-(v_-) + B_-(v_-)$ ; $K_4 = S_+F_-(v_+) - B_-(v_+)$ ;  (A.5)

$K_5 = S_-F_+(v_+) - B_+(v_+)$ ; $K_6 = S_-F_+(v_-) + B_+(v_-)$ ;

$K_7 = S_-F_+(v_-) + B_+(v_-)$ ; $K_8 = S_-F_+(v_+) - B_+(v_+)$ ;

$K_m = K_{m+8}$ ($m \geq 1$).

$$G_0 = \frac{K_{-1} + \Phi_0 K_{-2}}{F_+(v_-) + \Phi_0 F_+(v_+)} \; ; \; G_1 = \frac{K_1 + \Phi_1 K_2}{F_-(v_+) + \Phi_1 F_-(v_-)} \; ;$$

$$G_2 = \frac{K_3 + \Phi_1 K_4}{F_-(v_-) + \Phi_1 F_-(v_+)} \; ; \; G_3 = \frac{K_5 + \Phi_2 K_6}{F_+(v_+) + \Phi_2 F_+(v_-)} \; ;$$  (A.6)

$$G_4 = \frac{K_7 + \Phi_2 K_8}{F_+(v_-) + \Phi_2 F_+(v_+)} \; ; \; G_5 = \frac{K_9 + \Phi_3 K_{10}}{F_-(v_+) + \Phi_3 F_-(v_-)} \; ;$$

$$G_6 = \frac{K_{11} + \Phi_3 K_{12}}{F_-(v_-) + \Phi_3 F_-(v_+)} ; \quad G_7 = \frac{K_{13} + \Phi_4 K_{14}}{F_+(v_+) + \Phi_4 F_+(v_-)},$$

Where the factors $K_m$ can be obtained from the factors $P_m$ ( see Appendix 1) through the replacements:

$$S_+(v_\pm) \Leftrightarrow S_-(v_\pm); \quad F_+(v_\pm) \Leftrightarrow F_-(v_\pm); \quad B_+(v_\pm) \Leftrightarrow B_-(v_\pm). \tag{A.7}$$

**Figure captions.**

Figure 1 (color online): Multi-parameter gradient profiles of normalized potential U = $U(y)/U_0$, vs. normalized coordinate y = z/d. T he values of parameters in both graphs are M= 2.02, |g| = 1.35, L = 0.225 nm,. Fig. 1a – single barrier, Fig. 1b – single well; The dash-dotted lines represent the positions of the eigenstates calculated for these potentials in the text (for $U_0$= 1eV). The dashed line indicate the box-like potential with width y=1 and height equal to $U_{max} - U_{min}$ for curvilinear profile U(z). The dotted line shows the single energy state supported by this square well.

Figure 2 (color online): Different types of symmetric potentials (a) Double Well (DW) potential ($M_1 = M_2$ = 2.02, $|g_1|=|g_2|$=1.35, $L_1=L_2$=0.225 nm ). (b) (1) DW potential with up-shifted maximum ($M_1$ = 2.02, $|g_1|$ = 1.35, $L_1$ = 0.225 nm; $M_2$ = 2.843, $|g_2|$ = 0.971, $L_2$ = 0.44 nm ) and (2) DW potential with down-shifted maximum ( $M_1$ = 2.02, $|g_1|$ = 1.35, $L_1$ = 0.225 nm; $M_2$ = 4.739, $|g_2|$ = 2.036, $L_2$ = 0.35 nm ). The dashed and dash-dotted lines represent the levels associated with the of the eigenstates $\varepsilon_1$ and $\varepsilon_2$, respectively, evaluated for the potentials displayed.

Figure 3 (color online): Asymmetric DW potential ($M_1$ = 2.02, $|g_1|$ = 1.35, $L_1$ = 0.225 nm ; $M_2$ = 6.16, $|g_2|$ = 2.859, $L_2$ = 0.324 nm ).

Figure 4 (color online): Periodic multi-well potentials possessing (a) alternating convex ($M_1$ = 4.05, $g_1$ = -1.624, $L_1$ = 0.375 nm) and concave ($M_2$ =2.02, $g_2$ = 1.35, $L_2$ = 0.225nm) and (b) alternating convex ($M_1$ =2.02, $g_1$ = -1.35, $L_1$ = 0.225nm) and concave ($M_2$ = 4.05, $g_2$ = 1.624, $L_2$ = 0.375 nm) arcs.

Figure 5 (color online): Shape-dependent transmittance $|T|^2$ of single gradient barrier, shown on Fig. 1a (curve 1), and rectangular barrier (curve 2) with the same total width 2d and the same values of $U_{max}$ and $U_{min}$ for electrons with energy E plotted vs. normalized electron energy w = $E/U_{max}$ . According to the parameters of gradient barrier (Fig.1a) $U_{max}$ = 1.6 $U_0$, $p^2$ = 1/1.6. Thus, variable w corresponds to normalized electron energy $\varepsilon$, defined in (11), as w = $\varepsilon$/1.6.

Figure 6 (color online): Potential profiles 3a and 3b, obtained from profiles 1a and 1b, due to the transform (35), M = 2.02, |g| = 1.35. Profiles 1, 2 and 3 for single gradient barriers (Fig. 3a) and single wells (Fig. 3b) correspond to the values h = 1.225 ; 1 ; 0.8 respectively.

Figure 7 (color online): Normalized profile of potential barrier U= $U(y)/U_0$, y = z/d, providing the reflectionless tunneling of electron through the barrier with energy $\varepsilon$ in the range between $U_p$ and potential minimum $U_{min}$.

**Figure 1 :**

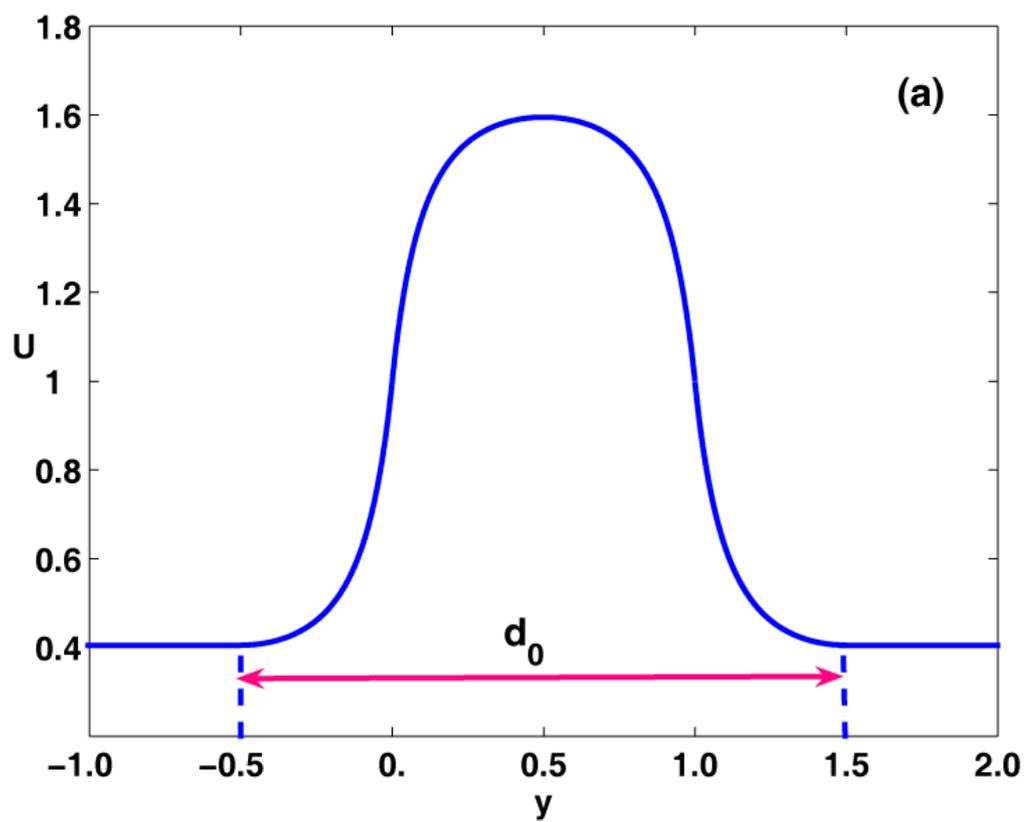

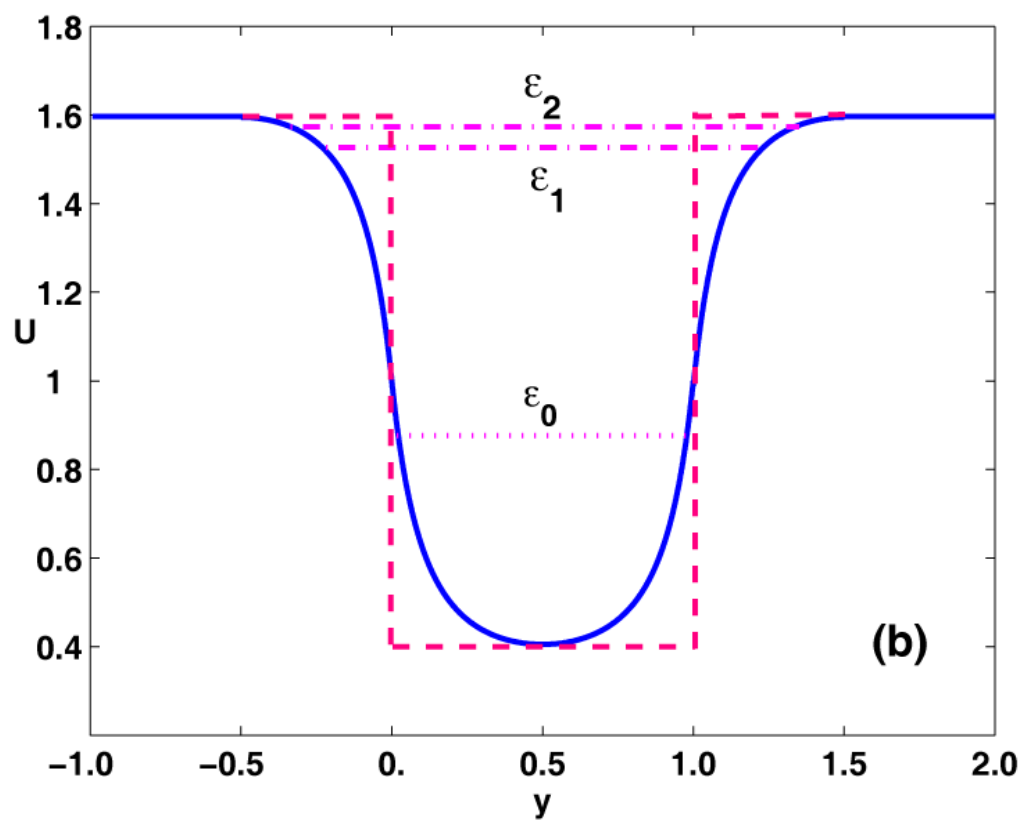

**Figure 2 :**

**(a)**

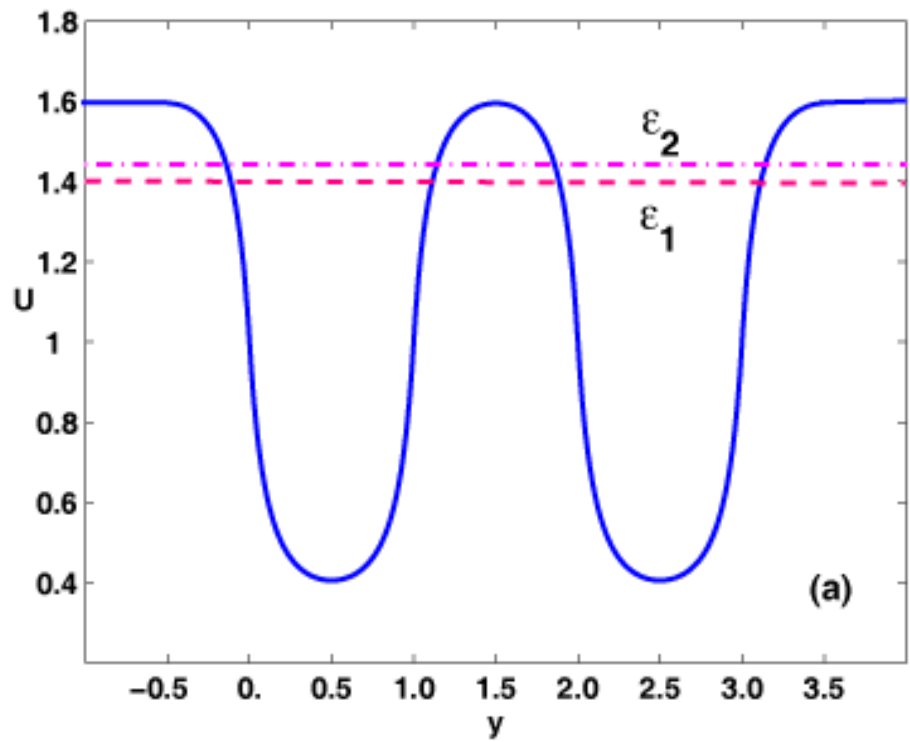

**(b)**

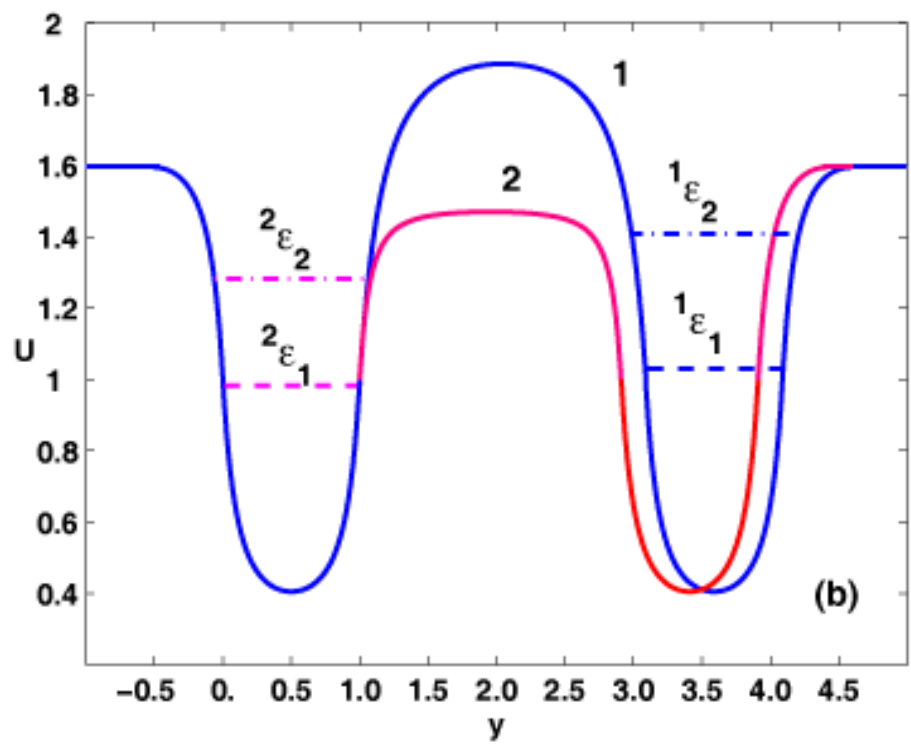

**Figure 3 :**

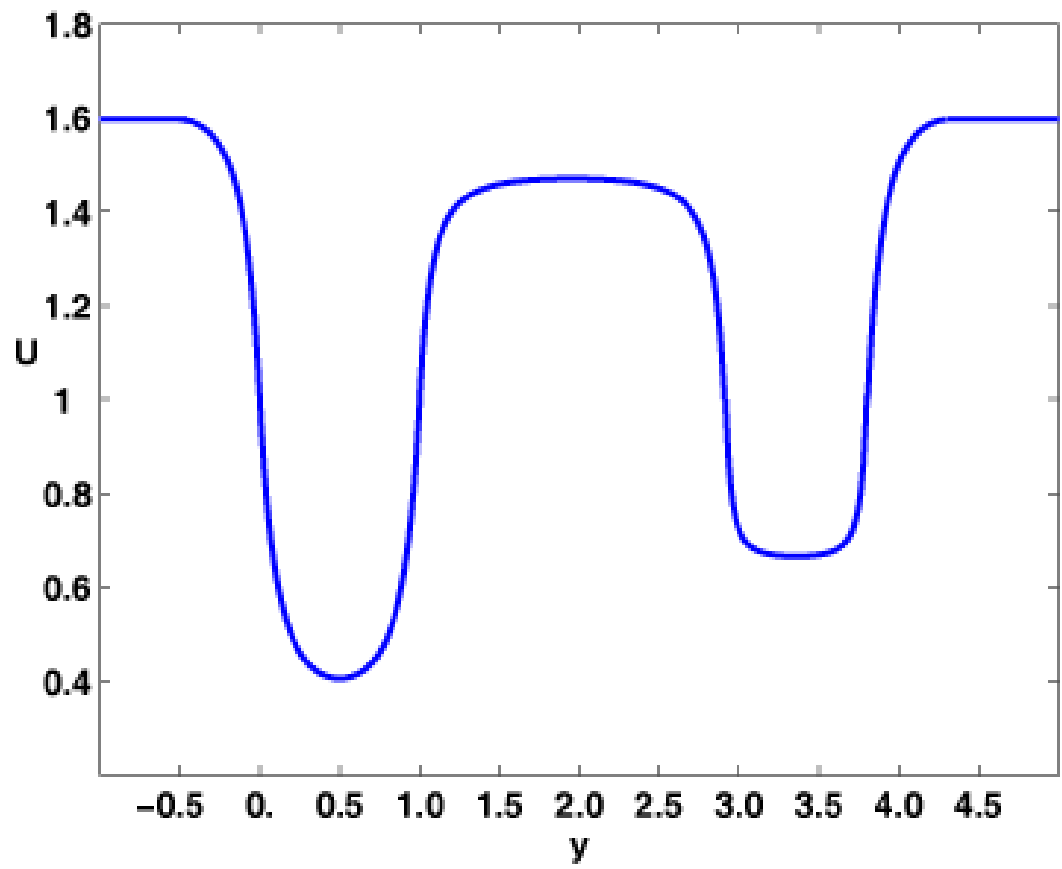

**Figure 4 :**

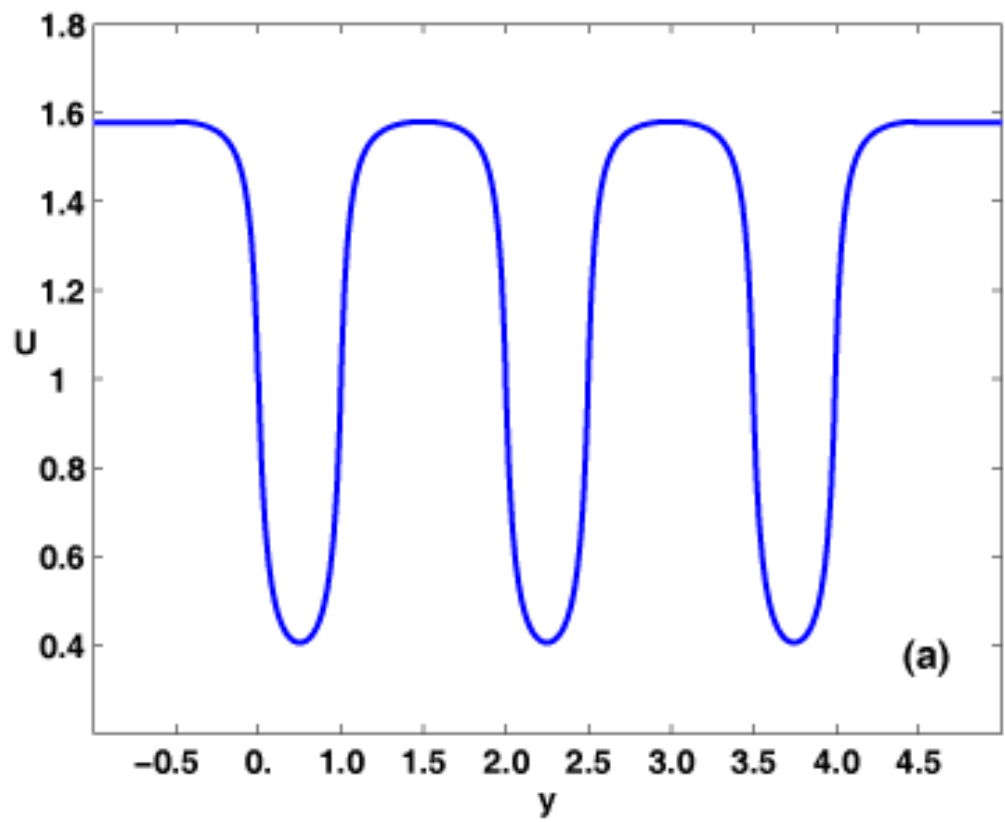

(a)

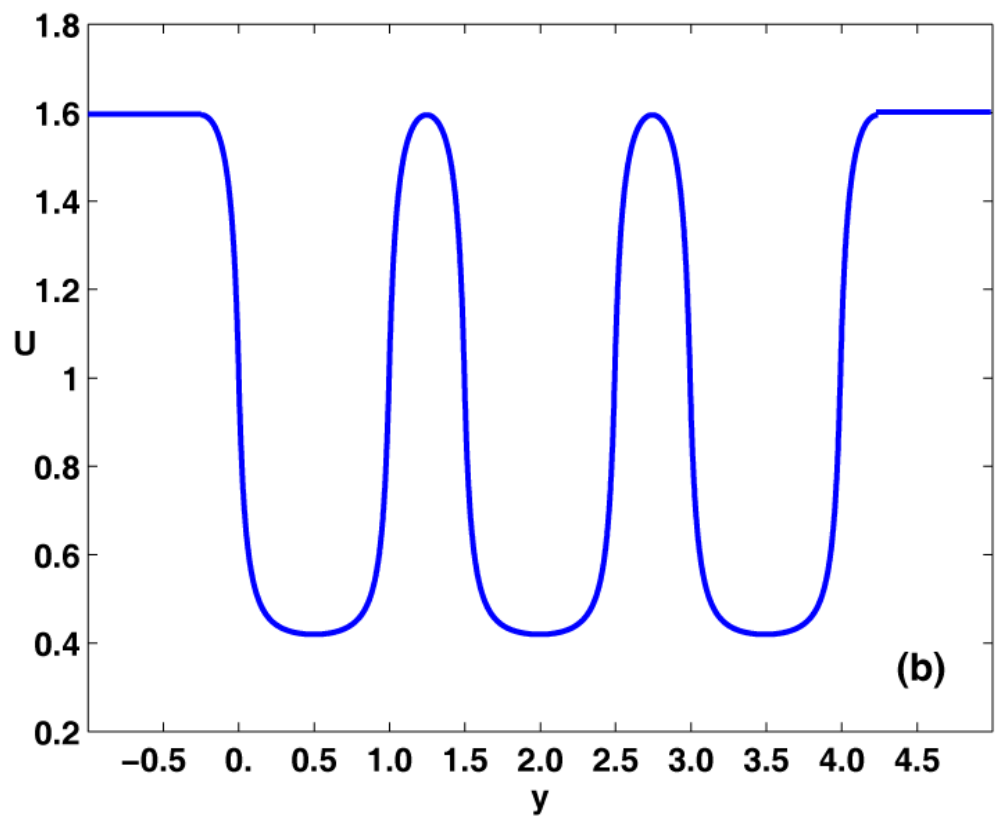

(b)

**Figure 5 :**

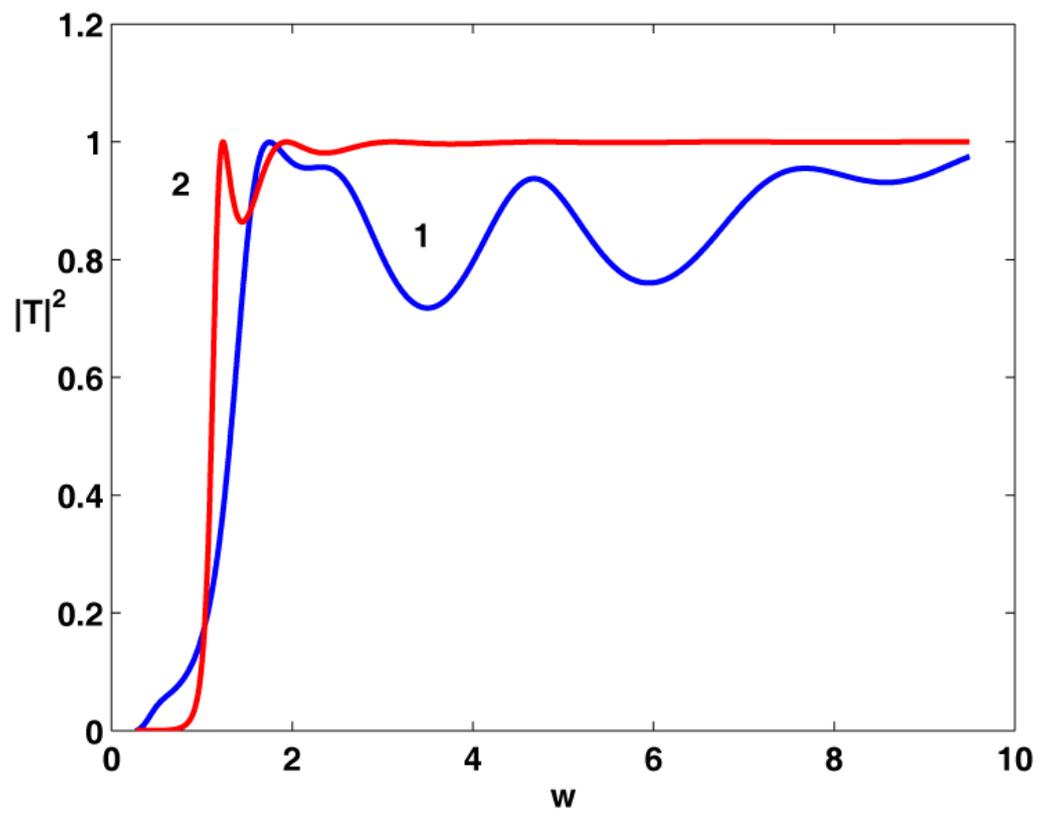

**Figure 6 :**

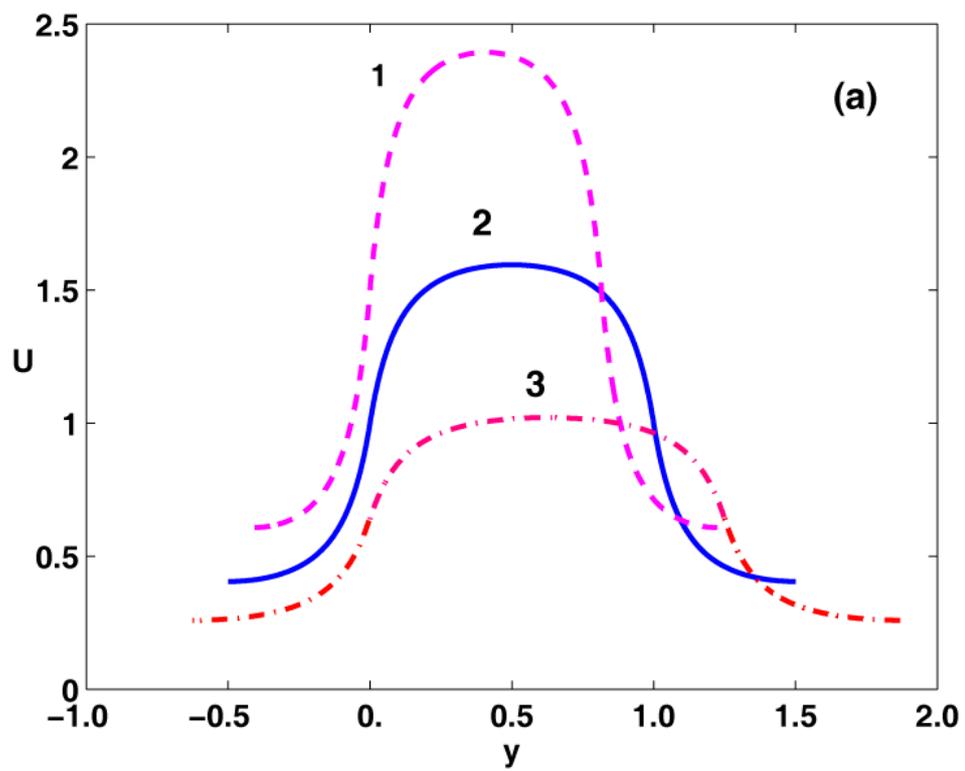

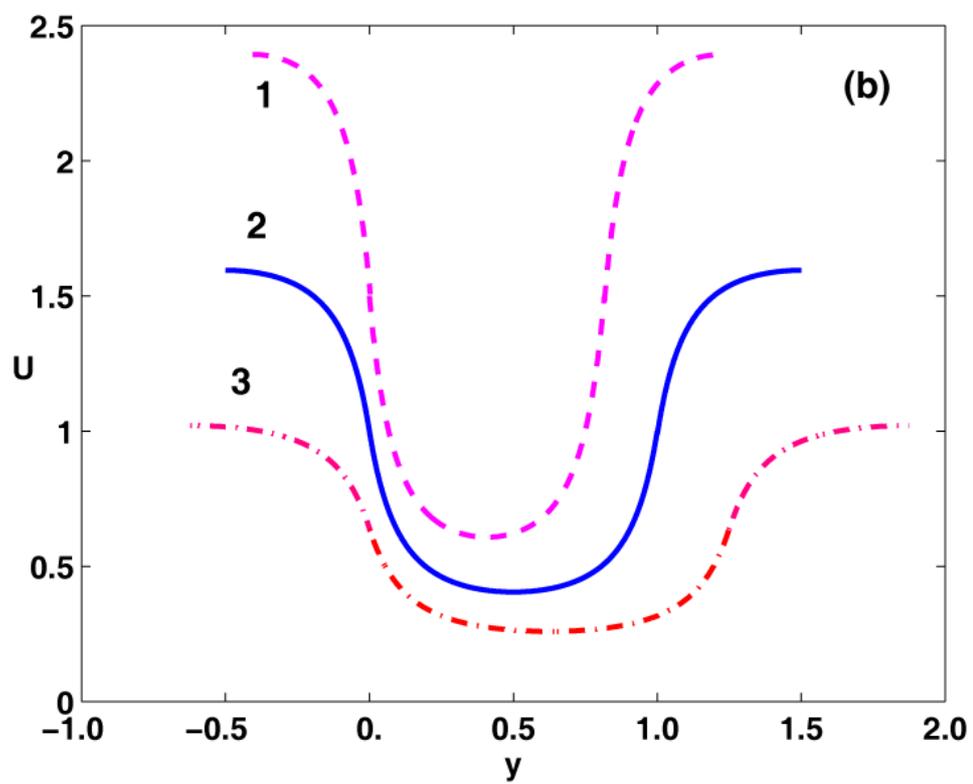

**Figure 7 :**

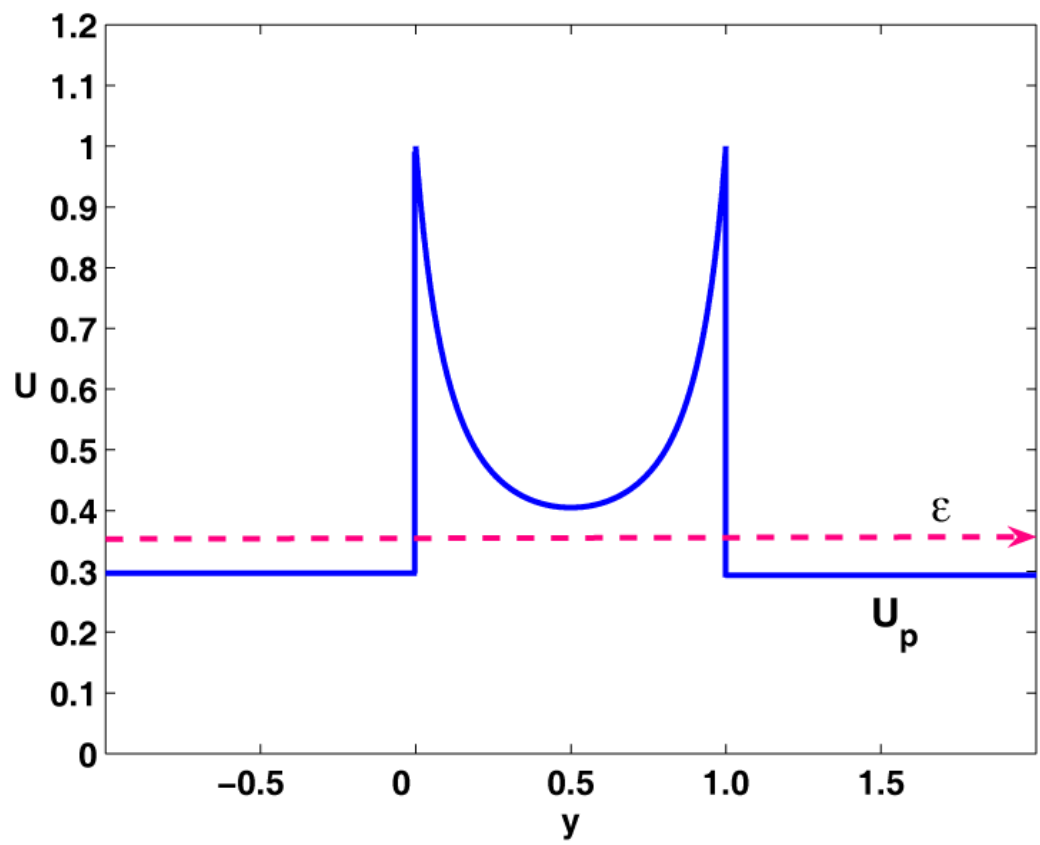